# Minimization of image watermarking side effects through subjective optimization

H. B. Golestani[1] and M. Ghanbari[1,2]

Sharif University of Technology, Tehran, Iran[1], and University of Essex, Colchester, UK[2]

h.b.golestani@gmail.com and ghan@essex.ac.uk

*Abstract*— **This paper investigates the use of Structural Similaritys (SSIM) index on the minimized side effect to image watermarking. For fast implementation and more compatibility with the standard DCT based codecs, watermark insertion is carried out on the DCT coefficients and hence a SSIM model for DCT based watermarking is developed. For faster implementation, the SSIM index is maximized over independent 4x4 non-overlapped blocks but the disparity between the adjacent blocks reduces the overall image quality. This problem is resolved through optimization of overlapped blocks, but, the higher image quality is achieved at a cost of high computational complexity. To reduce the computational complexity while preserving the good quality, optimization of semi-overlapped blocks is introduced. We show that while SSIM-based optimization over overlapped blocks has as high as 64 times the complexity of the 4x4 non-overlapped method, with semi-overlapped optimization the high quality of overlapped method is preserved only at a cost of less than 8 times the non-overlapped method.**

## I. Introduction

Digital watermarking is a technique of inserting a signature in a cover signal, such that the Human Visual System (HVS) is not able to recognize it, while, the inserted signature can be extracted using watermark detectors, or at least the existence of the watermark can be confirmed. Due to the widespread use of multimedia products, the watermarking technology is used in various applications. These include: prevention of unauthorized distribution [1], broadcast monitoring [2], multimedia authentication [3], inserting meta-data for error correcting or improving coding efficiency [4] and adding archived data to multimedia products [5] are some applications of digital watermarking.

In the watermarking design process, there is a trade-off between the three parameters of: robustness, imperceptibility and data payload. Robustness is the ability of watermark restoration after attacks. Imperceptibility, which is the main subject of this paper, addresses the quality degradation of watermarked image. Finally, after determining the acceptable requirements for robustness and imperceptibility; data-payload or storage capacity is defined as the number of bits that can be inserted into certain parts of the digital product. Depending on considered application of watermarking, a reasonable compromise among the above parameters should be maintained.

Watermarking systems try to insert data in such a way that the signal quality is preserved. This issue is crucial in some applications such as Medical Imaging [6]. However, controlling the side effect of inserting data on signal quality requires an objective metric for optimization. It is normally believed Mean Squared Error (MSE) objective quality is not a good measure for such a case, since it does not well correlate to the subjective quality measures [7]. In the past decade the Video Quality Experts Group (VQEG) [8] have recommended a set of quality metrics, with or without the reference to non-distorted images or even to partial references, where the objective image quality well correlates to the subjective measures. However, these tools are most suited for measurements of processed images, rather than using them to control the amount of inserted distortions on the fly. Perhaps measure of structural similarity index (SSIM) [9] is a much better and easier way of online control of subjective image degradation.

After the introduction of the SSIM index in 2004, a widespread use of it for evaluating the quality of image and video processing systems began. It was even used in optimization of visual systems such as: Rate-Distortion (RD) optimization in video coding [10], image quantization [11], image compression [12], image coding [13], image de-noising [14], motion estimation optimization [15] and contrast enhancement [16].

SSIM index has also been used in digital watermarking [16] [17]. In [16] the global histogram manipulation of original image is controlled with the SSIM index of overlapping Gaussian windows. However, as we shall see, this type of windowing drastically increases the computational complexity in the optimization process. They have solved this complex problem through a time consuming iterative method which still obtains a sub-optimum solution. [17] addresses the problem of selecting the best set of DCT coefficients from the whole, where the whole image is transformed into DCT domain and $2N$ mid-frequency coefficients are selected by default. To select the $N$ best coefficients in terms of the SSIM index, the Particle Swarm Optimization (PSO) is incorporated with SSIM index as a fitness function. The high computational complexity and the required access to the original image to extract the watermark, are the two major drawbacks of this system.

In this paper we aim to minimize the side effect of image watermarking on the standard DCT based codecs. Since the SSIM index is defined in the spatial domain, SSIM-based DCT domain watermarking can not be modeled directly. In this paper, a 2D DCT domain SSIM Index is introduced and employed to provide a SSIM index model for the DCT domain watermarking. The major advantage of this model is that optimized watermarking parameters can be determined directly in the DCT domain and there is no need to carry out some time-consuming decoding processes.

To simplify the optimization complexity and speed-up the embedding procedure, the SSIM index is used in "non-overlapped" blocks. That is, the optimization parameters are determined such that the SSIM index of non-overlapped square blocks is maximized. The results show that the optimization in non-overlapped mode can preserve the quality of watermarked image in the low to medium watermarking strengths. To improve robustness, one may increase watermarking strength but it causes some disparity between the non-overlapped blocks. In the following, the source of disparity side-effect is analyzed and some solutions are suggested.

The first solution is optimization over the "overlapped" blocks. In this method, the surrounding neighboring blocks of the block under watermark insertion are used in the optimization process, such that the visual quality of the watermarked signal is greatly improved. A major disadvantage of the overlapped method is its too high computational complexity. Perhaps the proposed optimization of "semi-overlapped" blocks is the best solution to overcome the problem. This approach not only preserves the original quality of the image as good as the overlapped method but it also has much less computational complexity and provides a blind watermark extraction.

The rest of the paper is organized as follows. In Section II, the spatial domain SSIM index is studied. The DCT domain SSIM index for the design of SSIM-based image watermarking and its requirements are addressed in section III. Section VI is devoted to discussions and analysis of the preliminary results. The overlapped and the proposed semi-overlapped optimization methods are presented in section V and concluding remarks are given in section VI.

## II. Structural Similarity Index (SSIM Index)

Pixels of natural images are highly correlated to each other, and these correlations provide information about objects and structures in the image. The SSIM index is a full-reference quality meter that extracts texture and structural information of both original and processed images for comparison. Luminance, contrast and structural similarity are three independent characteristics that could be compared between the two images. The SSIM index between two images $x$ and $y$ is defined as [9]:

$$SSIM(\boldsymbol{x},\boldsymbol{y}) = l(\boldsymbol{x},\boldsymbol{y})^{\alpha}\, c(\boldsymbol{x},\boldsymbol{y})^{\beta}\, s(\boldsymbol{x},\boldsymbol{y})^{\gamma} \qquad (1)$$

where $l(\boldsymbol{x},\boldsymbol{y})=(2\mu_x\mu_y+C_1)/(\mu_x^2+\mu_y^2+C_1)$ compares the luminance likeness, $c(\boldsymbol{x},\boldsymbol{y})=(2\sigma_x\sigma_y+C_2)/(\sigma_x^2+\sigma_y^2+C_2)$ evaluates contrast similarity and $s(x,y)=(\sigma_{xy}+C_3)/(\sigma_x\sigma_y+C_3)$ measures structural correlation between $x$ and $y$. Quantities $\mu_x$, $\sigma_x$ are the mean and variance of $x$, respectively and $\sigma_{xy}$ is the sample cross-covariance between $x$ and $y$. Constants $C_1$, $C_2$ and $C_3$ are defined to prevent ambiguous condition 0/0 in dark or smooth regions of $x$ and $y$. Also, parameters $\alpha>0$, $\beta>0$ and $\gamma>0$ adjust the weight of $l(x,y)$, $c(x,y)$ and $s(x,y)$ respectively. For $\alpha = \beta = \gamma = 1$ and $C_3 = C_2 / 2$, (1) is simplified to (2) with $C_1$, $C_2$ are derived from $C_1=(K_1L)^2$ and $C_2=(K_2L)^2$. For 8-bit images, L is 255, also $K_1$ and $K_2\ll1$ are very small constants.

$$SSIM(\boldsymbol{x},\boldsymbol{y})=\left(\frac{2\mu_x\mu_y+C_1}{\mu_x^2+\mu_y^2+C_1}\right)\left(\frac{2\sigma_{xy}+C_2}{\sigma_x^2+\sigma_y^2+C_2}\right) \qquad (2)$$

To evaluate image quality, it is better to apply SSIM index locally rather than globally [9]. To determine the local SSIM index, the local statistics $\mu_x$, $\sigma_x$ and $\sigma_{xy}$ are computed for a specified window and then the local SSIM index is computed from (2). The sliding window moves to cover the entire image and the overall SSIM index is derived from the average of local SSIM indices. Here we define "overlapped" mode when local windows do overlap, and "non-overlapped" mode when the local windows are separate. The SSIM index in the overlapped mode is more accurate than the non-overlapped mode, but its high computational complexity has led many researchers to use the non-overlapped mode in their optimization [10]-[12]-[14]-[15]. However, the Pioneers of SSIM index suggest an 11×11 circular-symmetric Gaussian window (called *Gaussian window*) to be used in the overlapped mode. Note that the SSIM index is a number between +1 and -1 and being closer to +1 indicates higher quality of the processed image.

## III. SSIM Based Watermarking

Watermarking can be carried out in both spatial and frequency domains, but frequency domain watermarking is preferred [18]. Among all the frequency transforms, the DCT is the most widely used transform due to its appropriate features hence standard image and video compression techniques use DCT is their frequency transform.

Unfortunately, presented SSIM index in [9] uses spatial samples and it is not usable in the DCT domain directly. In 2008, Channappayya et. al. transformed the conventional SSIM index into 1D DCT domain to find lower and upper bounds for the SSIM index of quantized images [19]. With a similar method, we present a 2D version of it in the following form:

$$SSIM(\boldsymbol{x},\boldsymbol{y})=\left(\frac{2\boldsymbol{X}_{00}\boldsymbol{Y}_{00}+C_1'}{\boldsymbol{X}_{00}^2+\boldsymbol{Y}_{00}^2+C_1'}\right)\times$$
$$\left(\frac{2\sum\limits_{p=0}^{N-1}\sum\limits_{q=0}^{N-1}(\boldsymbol{X}_{pq}\otimes\boldsymbol{Y}_{pq})-2\boldsymbol{X}_{00}\boldsymbol{Y}_{00}+C_2'}{\sum\limits_{p=0}^{N-1}\sum\limits_{q=0}^{N-1}(\boldsymbol{X}_{pq}\otimes\boldsymbol{X}_{pq}+\boldsymbol{Y}_{pq}\otimes\boldsymbol{Y}_{pq})-\boldsymbol{X}_{00}^2-\boldsymbol{Y}_{00}^2+C_2'}\right) \qquad (3)$$

where $N\times N$ matrices $\boldsymbol{X}$,$\boldsymbol{Y}$ are the DCT domain representation of pixel matrices $x$,$y$. $\boldsymbol{X}_{pq}$ and $\boldsymbol{X}_{00}$ stand for $(p+1,q+1)$ coefficient and the DC coefficient, respectively. $\otimes$ is the element-by-element product and constants $C_1'$ and $C_2'$ are computed by $C_1'=N^2C_1$ and $C_2'=(N^2-1)C_2$. The first term evaluates luminance correlation and only uses DC coefficients. The last term is used to compare contrast and structural similarity and uses AC Coefficients. Since no approximation is used in the SSIM index translation from spatial domain to DCT domain, then the derived SSIM index from (2) and (3) are exactly the same.

Equation (3) is in fact what is needed to calculate SSIM index in the DCT domain. We will use it to study a SSIM-based watermarking scheme in three steps: problem modeling, embedding pattern design and calculating optimum watermarking parameters. These three steps are explained in the following sub-sections.

### A. Modeling DCT domain Watermarking

In this section, DCT domain watermarking is first modeled in view of SSIM index and then a set of appropriate coefficients are selected for embedding. Assume $x$, $y$ are two 4×4 original and watermarked image blocks, respectively (in line with block size in H.264 video codec). The SSIM index between $x$, $y$ is obtained by setting $N = 4$ in (3). Suppose for embedding a bit into $X$, $W$ must be added to $X$ where $Y=X+W$ is the watermarked block. Since the Human Visual System (HVS) is very sensitive to DC coefficient degradation, this coefficient is not altered ($W_{00}=0$) so $X_{00}=Y_{00}$ and the first term in (3) is equal to 1 so it has no effect on our modeling. With the assumption, $A_X \triangleq X_{00}{}^2 + X_{01}{}^2 + \dots + X_{33}{}^2$, $C_X \triangleq A_X - 2X_{00}{}^2 + C_2'$ is a constant independent of $W$. We have:

$$SSIM(x, y) = \frac{2 \sum_{p=0}^{3} \sum_{q=0}^{3} X \otimes W + C_X}{\sum_{p=0}^{3} \sum_{q=0}^{3} W \otimes W + 2 \sum_{p=0}^{3} \sum_{q=0}^{3} X \otimes W + C_X}$$

$$= 1 - \frac{\sum_{p=0}^{3} \sum_{q=0}^{3} W \otimes W}{\sum_{p=0}^{3} \sum_{q=0}^{3} W \otimes W + 2 \sum_{p=0}^{3} \sum_{q=0}^{3} X \otimes W + C_X} \qquad (4)$$

Equation (4) shows that in addition to the additive watermark $W$, the original DCT coefficients $X$ and specifically quantity $\Sigma_{p=0}^{3}\Sigma_{q=0}^{3}W \otimes X$ are also determinate. Since the main purpose of the paper is to minimize the side effect of watermarking through optimization and not designing an efficient watermarking method, then a simple method for watermarking is used.

According to the above discussion, as an example we chose two low frequency coefficients $X_{01}$ and $X_{10}$ for embedding. If due to embedding these coefficients are altered by $\varepsilon$ and $\sigma$ respectively, the SSIM index due to this introduced error derived from (4) is simplified to (5). Throughout the paper (5) is used in our investigation as a similarity measure. The watermarking method should reduce the second term in (5) to result in a better visual quality and this is obtained by selecting appropriate $\varepsilon$ and $\sigma$ values based on watermarking rule (embedding pattern) and image charachteristics of $X_{01}$, $X_{10}$ and $C_X$.

$$SSIM(x, y) = 1 - \frac{\varepsilon^2 + \sigma^2}{\varepsilon^2 + \sigma^2 + 2\varepsilon X_{01} + 2\sigma X_{10} + C_X} \qquad (5)$$

### B.  Simple Embedding Pattern

A simple type of Quantization Index Modulation (QIM) embedding scheme, which provides good rate-distortion-robustness performance [20], is employed in our study. The original coefficients $X_{01}$ and $X_{10}$ are modified as $Y_{01} = X_{01} + \varepsilon$, $Y_{10} = X_{10} + \sigma$ according to (6) where $k$ is an integer and $S$ controls the watermarking strength and visibility distortion . For blind extraction of the watermarked "bit" one may subtract the two watermarked coefficient $Y_{12}$ and $Y_{21}$ according to (7).

$$Y_{12} - Y_{21} = (X_{12} + \varepsilon) - (X_{21} + \sigma) = \begin{cases} (2k + \frac{1}{2})S & bit = 1 \\ (2k - \frac{1}{2})S & bit = 0 \end{cases} \qquad (6)$$

$$bit = \begin{cases} 0 & \mathrm{mod}(Y_{12} - Y_{21}, 2S) \leq S \\ 1 & \mathrm{mod}(Y_{12} - Y_{21}, 2S) > S \end{cases} \qquad (7)$$

### C.  Optimum Watermarking Parameters

In this section, according to the model presented in (5) and the watermarking rule presented in (6) the optimal values of $k$, $\varepsilon$ and $\sigma$ will be determined. Suppose bit "1" is to be inserted. In this case, (6) becomes:

$$(X_{01} + \varepsilon) - (X_{10} + \sigma) = (2k + \frac{1}{2})S \qquad (8)$$

Determining $\sigma$ in terms of the others, results in:

$$\sigma = \varepsilon - (2k + \frac{1}{2})S + X_{01} - X_{10} \qquad (9)$$

Substituting (9) in (5) results in the SSIM index in terms of $\varepsilon$ and $k$ leading to  (10)

$$SSIM(x, y) = 1 - \frac{\varepsilon^2 + (\varepsilon - (2k + \frac{1}{2})S + X_{01} - X_{10})^2}{\varepsilon^2 + (\varepsilon - (2k + \frac{1}{2})S + X_{01} - X_{10})^2 + 2\varepsilon X_{01} + 2(\varepsilon - (2k + \frac{1}{2})S + X_{01} - X_{10})X_{10} + C_X} \qquad (10)$$

Parameters $\varepsilon \in$ R and $k \in$ Z must be determined in such a way that the SSIM index is maximized or equivalently the second term of (10) is minimized. Presenting an analytical solution for maximizing (10) is difficult but as a rule of thumb, the expression in the numerator of the second term in (10) should be minimized and the expression $2\varepsilon X_{01}+2(\varepsilon-(2k+0.5)S+X_{01}-X_{01})X_{10}$ in its denominator should be maximized. Fig. 1 shows a 3D SSIM index function for a $4 \times 4$ block of Lena image in terms of $\varepsilon$ and $k$. The maximum point of the SSIM index usually occurs for non-integer values of $k$ but due to the watermarking rule (6), $k$ must be an integer value. Hence we can approximate the peak value to one of its surrounding points which has integer $k$ value. As shown in Fig. 1, moving away from the maximum point reduces the SSIM index but to maximize image quality for an integer $k$, we may move in the direction that the SSIM index has the smallest loss. For Fig. 1, the largest reachable SSIM index is 95.90% and it is obtained at $k = 0$ and $\varepsilon = 6.69$. The next highest SSIM index is held at $k = -1$ and $\varepsilon = -26.42$ at the value of 81.36%. Finally $\sigma$ is calculated from (9) and the pairs $(\varepsilon, \sigma)$ will be at their optimum values. Note that the contour lines plotted in the $k$-$\varepsilon$ plain can be employed to anticipate optimal values of $\varepsilon$ and $k$.

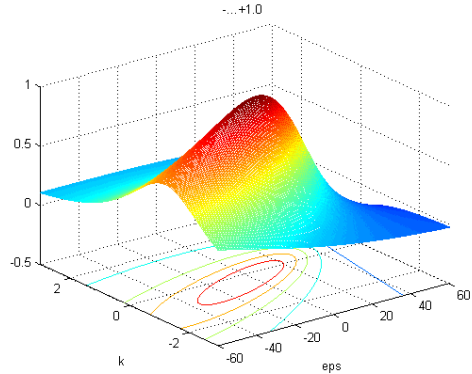

Fig. 1.  A 3D plot of SSIM index in terms of $\varepsilon$ and $k$

## IV. Preliminary Results and Discussions

To evaluate our method on various image textures, a medium detailed part of the Lena image, a high detailed section of the Baboon image and finally a low detail part of the Peppers image, are selected as original images. These 24x24 pixels images are then decomposed into 36 non-overlapping $4\times4$ blocks and a random bit is inserted into each block. In this section, the minimization of image watermarking side effects is evaluated for various power strengths of $S$. The original images, the watermarked images and the computed SSIM indices are depicted in Figs. 2-4.

Two important notes can be seen from Figs. 2-4. First, increasing the watermarking strength $S$ reduces image quality, which is natural, as well as reducing the average SSIM index. This is because the larger the values of $S$ the sharper becomes the 3D SSIM index function, as seen from Fig 5, hence increasing the sensitivity of SSIM to the optimised parameters of $k$ and $\varepsilon$. Considering that $k$ needs to be an integer value, sharper decay of SSIM for search of an integer $k$, will greatly devite SSIM from its maximum values. The second note is that highly detailed images provide larger range for watermarking strength S than lower detailed images, for the same visual quality. Comparing Fig. 3d and Fig.4d verify this, since they have the same SSIM indices but the image with more details (Fig.3) uses S=120 while the image with low detail uses S=45.

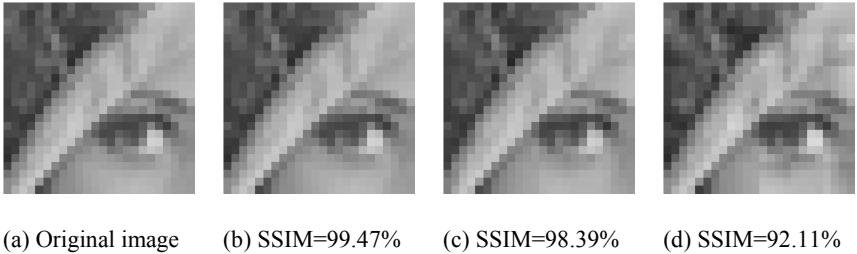

(a) Original image     (b) SSIM=99.47%     (c) SSIM=98.39%     (d) SSIM=92.11%

Fig. 2. Quality of watermarked images at various watermark strengths $S$. (a) original non-watermarked image, (b)-(d) watermakred images with: (b) $S$=15 (c) $S$=30 (d) $S$=60.

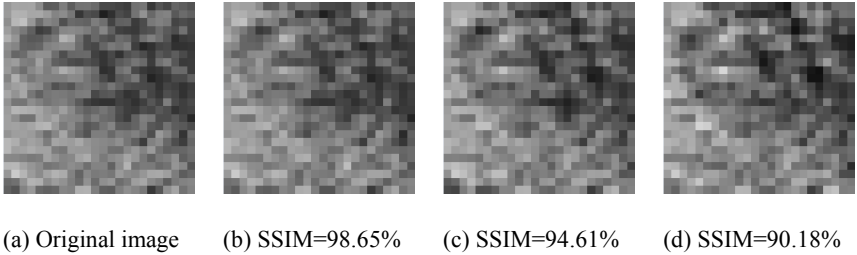

(a) Original image     (b) SSIM=98.65%     (c) SSIM=94.61%     (d) SSIM=90.18%

Fig. 3. Quality of watermarked images at various watermark strengths $S$. (a) original non-watermarked image, (b) –(d) watermakred images with: (b) $S$=40 (c) $S$=80 (d) $S$=120.

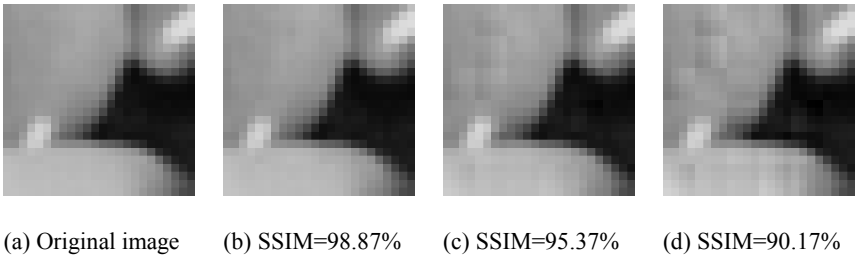

(a) Original image     (b) SSIM=98.87%     (c) SSIM=95.37%     (d) SSIM=90.17%

Fig. 4. Quality of watermarked images at various watermark strengths $S$. (a) original non-watermarked image, (b)-(d) watermakred images with: (b) $S$=15 (c) $S$=30 (d) $S$=45.

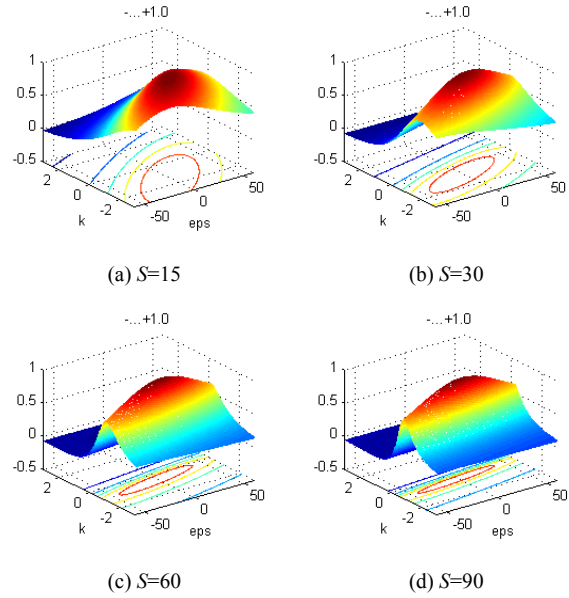

(a) $S$=15             (b) $S$=30

(c) $S$=60             (d) $S$=90

Fig. 5. Sensitivity of SSIM index to strength parameter $S$

The proposed SSIM-based method is based on the maximization of SSIM of non-overlapping blocks. One may increase $S$ to provide higher robustness. At larger values of $S$, individual optimization of SSIM may create a large disparity between the adjacent blocks, reducing the overall image SSIM as well as its perceptual quality. Figure 6 shows how optimized SSIM at larger values of $S$ degrades the overall perceptual quality of the image, while its side effect at smaller values of $S$ is negligible.

In the next section, we first deal with the disparity problem and then through introduction of "optimization of the semi-overlapped blocks" we aim to improve the watermarked image quality.

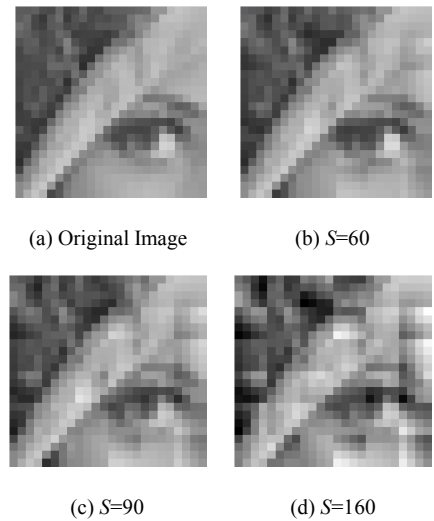

(a) Original Image        (b) $S$=60

(c) $S$=90            (d) $S$=160

Fig. 6. Independently optimized watermarked blocks, (a) original image, and (b)-(d) watermarked images with:(b) $S$=60, (c) $S$=90, (d) $S$=160.

## V. Solutions for Removing the Disparity Effect

To overcome the disparity side effect, neighbors of each block should be used in the optimization process. Hence optimization in the overlapped blocks may overcome this disparity.

### A. Optimization of Overlapped Blocks

Consider an $R \times S$ size original image. This image is decomposed into $RS/16$ non-overlapped $4 \times 4$ blocks and due to (6) a pair of $(\varepsilon^{pq}, \sigma^{pq})$ or $(\varepsilon^{pq}, k^{pq})$ are assigned to each block ($1 \le p, q \le RS/16$). An $N \times N$ pixel sliding window moves pixel-by-pixel and the local SSIM indices are derived in terms of spesific $\varepsilon^{pq}$ and $k^{pq}$. For example, the red window in Fig. 7a includes $(\varepsilon^{11}, k^{11})$ and $(\varepsilon^{12}, k^{12})$. The optimization runs for each local sliding window and its optimum parameters are calculated. The averages of the derived parameters become the final values of the pair for the block.

The main advantage of optimization in the overlapped mode is that it covers all the edges around the current $4 \times 4$ block and removes any disparity between it and its neighbors. This is achieved at a cost of high computational complexity.

The computational complexity is measured through $C_{over.} = A_{over.} \times B_{over.}$. $A_{over.}$ is defined as the number of optimizations carried out per image and is equal to the number of different sliding windows covering the entire image. For an $R \times S$ size image and $N \times N$ non-weighted sliding window size, $A_{over.}$ is given in (11). $N$ cannot be greater than the image size and it should be equal or greater than 4 ($N \ge 4$) otherwise some intra-block optimization processes which does not cover any edges, are carried out. If an $N \times N$ weighted sliding window (e.g. Gaussian window suggested by pioneers of SSIM Index [9]) is employed, first the original image should be zero-padded. In this case $A_{over.}$ is a bit greater than (11) and the approximation can be neglected in normal image sizes.

$$A_{over.} = (R - N + 1)(S - N + 1), \quad 4 \le N \le \min(R, S) \qquad (11)$$

$B_{over.}$ is defined as the number of performed operations per optimization and is an exponential function of the number of blocks seen through the sliding window. For example, only one block can be seen through the black sliding window in Fig. 7a, though the red one contains two blocks. If $N$ is a power of 2, the approximate value of $B_{over.}$ is given in (12). When $N$ is incremented $A_{over.}$ is reduced but $B_{over.}$ is increased exponentially such that $C_{over}$ is increased too. Thus in view of the computational complexity smaller sliding windows is preferred.

$$B_{over.} \simeq \frac{1}{16}(2^{(N/2)^2} + 6 \times 2^{(N/2)(N/2+1)} + 9 \times 2^{(N/2+1)^2}) \qquad (12)$$

Another drawback of using wider sliding windows is the inaccuracy of the SSIM index. To investigate the impact of sliding window size on the accuracy of the SSIM index we have calculated the SSIM index of the TID2008 database [21] for a range of window sizes, and the results are shown in Fig 8. This database contains 1700 distorted images of various kinds with given MOS (Mean Opinion Score) values. Fig. 8 illustrates the scatter plot of the scaled MOS values and the calculated SSIM index through both 11×11 Guassian window suggested by [9] and various sizes of square shape overlapped windows. The closeness of the scatter points to the straight line $y = x$, indicates a higher correlation between the calculated SSIM index and the MOS values. As Fig 8 shows, while the overlapped small 4x4 square window behaves like the 11×11 Guassian one, at larger overlapped square windows SSIM values are shifted to higher values, inflating their values . Larger departures between the 11×11 Guassian overlapped and square overlapped windows of various sizes are shown in Table I, and hence 4x4 and 8x8 square sliding windows seem to be a good choice, in addition to be suited for H.264 video codecs.

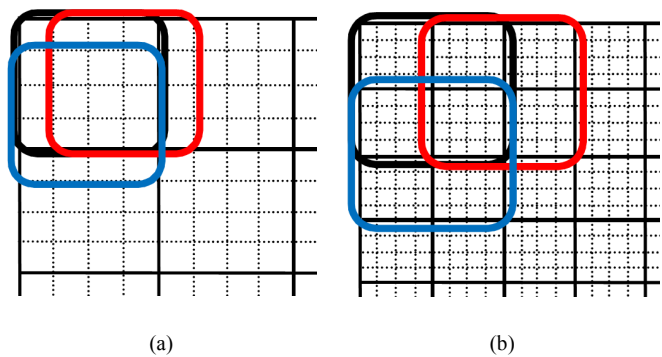

(a)                    (b)

Fig. 7.   Sliding windows, (a) 4x4 overlapped, (b) 8×8 semi-overlapped

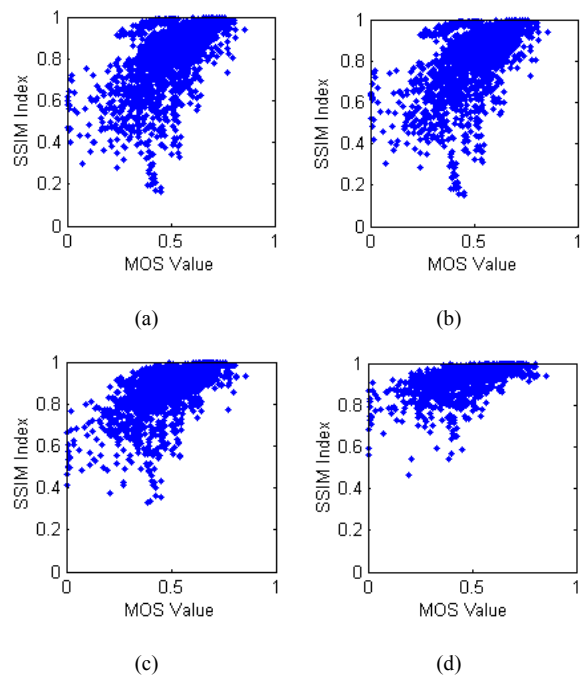

Fig. 8. Scatter plot of SSIM index vs MOS, for (a) the Gaussian overlapped window, (b)-(d) non-overlapped windows of size: (b) 4x4, (c) 32x32, (d) 128x128.

TABLE I

$L^2$-norm of difference between the SSIM indices and the MOS values

| Window | Gaussian | 4×4 | 8×8 | 16×16 | 32×32 | 64×64 |
|---|---|---|---|---|---|---|
| $L^2$-norm of error | 189 | 186 | 193 | 221 | 259 | 305 |

*B.  Optimization of Semi-overlapped Blocks*

In the semi-overlapped mode, an 8x8 sliding window scans the entire image with 4-pixel length steps, as shown in Fig. 7b. In this mode, the objective is to find optimum watermarking parameters $\varepsilon$ and $\sigma$ of each block based on optimization of the SSIM index of an 8×8 semi-overlapped sliding window. As before, the average values of the derived parameters of the current block become the final values of that block.

Similar to optimization of the overlapped blocks, the computational complexity of semi-overlapped method is measured as $C_{semi.} = A_{semi.} \times B_{semi.}$ where, for a $R×S$ image size, $A_{semi.}$ is given in (13) and since each sliding window covers 4 blocks, $B_{semi.}$ is equal to $2^4$. Normally, not only $A_{semi.}$ is about 16 times less than $A_{over.}$ but also $B_{semi.} << B_{over.}$ so the computational complexity of semi-overlapped method is much less than that of overlapped one.

$$A_{semi.} = (R/4 - 1) \times (S/4 - 1) \qquad (13)$$

In the following, a complete comparison of the non-overlapped, overlapped with 4×4 and Gaussian windows and the semi-overlapped method is presented. The original and watermarked images with strength $S$=160 for a part of the Lena (medium detail), S=90 for a part of the Peppers image (low detail) and S=400 for a part of the Baboon image (high detail) are presented in Figs. 9-11, respectively. As can be seen, the semi-overlapped method preserves the subjective quality of the watermarked image as good as both overlapped methods (4x4 and Gaussian), however its computational complexity is much less than the two. We have also verified these claims through subjective tests, where on the average subjects could not discriminate between Figs 9-11 (c-e), whereas they could clearly differenciated them from Figs 9-11 (b).

Comparison of computational complexity for CIF image format (360×288) is reported in Table II. The first raw of this table is the number of optimizations carried out per image and the second raw is the number of performed operations per optimization and the $3^{rd}$ raw indicates the total carried out operations per image. For an easy comparison, the $3^{rd}$ raw enteries are normalized to that of the non-overlapped value and the results are presented in the $4^{th}$ raw. These results demonstrate that the computational complexity of the semi-overlapped method is much less than the overlapped ones (4x4 and Gaussian).

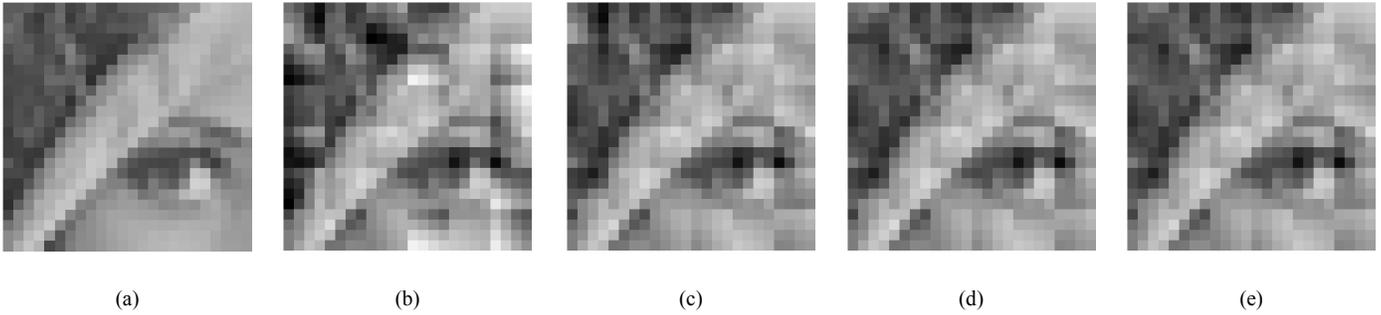

(a)  (b)  (c)  (d)  (e)

Fig. 9. Comparison of presented methods for watermarking strength S=160, (a) original medium-detailed image, with watermarked images using (b) non-overlapped, (c) overlapped with 4×4 sliding windows, (d) overlapped with 11×11 standard Gaussian windows and (e) semi-overlapped

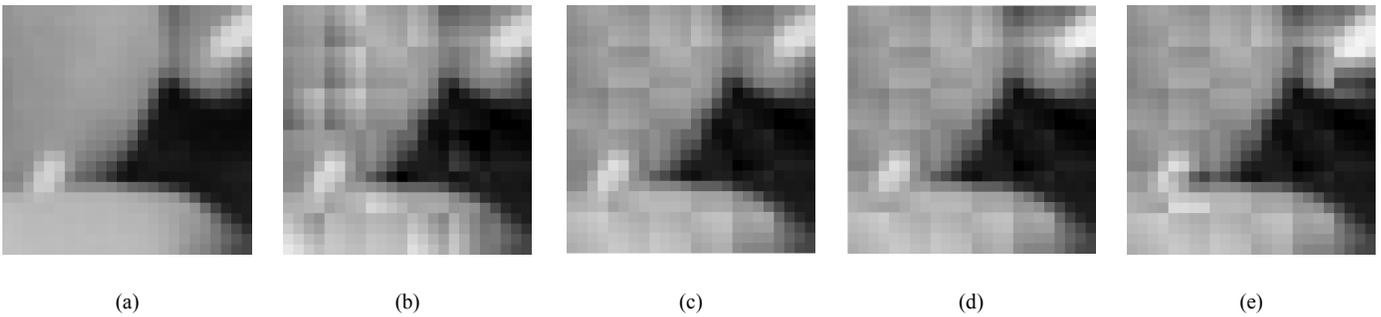

(a)  (b)  (c)  (d)  (e)

Fig. 10. Comparison of presented methods for watermarking strength S=90, (a) original low-detailed image, with watermarked images using (b) non-overlapped, (c) overlapped with 4×4 sliding windows, (d) overlapped with 11×11 standard Gaussian windows and (e) semi-overlapped

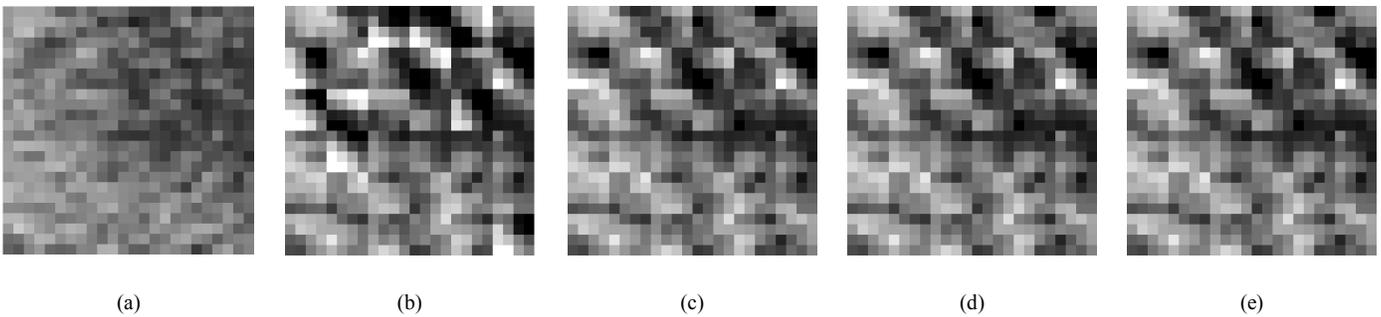

(a)  (b)  (c)  (d)  (e)

Fig. 11. Comparison of presented methods for watermarking strength S=400, (a) original high-detailed image, with watermarked images using (b) non-overlapped, (c) overlapped with 4×4 sliding windows, (d) overlapped with 11×11 standard Gaussian windows and (e) semi-overlapped

TABLE II

COMPARISON OF COMPUTATIONAL COMPLEXITY OF THE PRESENTED METHODS FOR CIF IMAGE FORMAT (360×288)

| | 4×4 Non-overlapped | 4x4 overlapped | Gaussian overlapped | 8×8 Semi-overlapped |
|---|---|---|---|---|
| # optimizations per image | 6480 | 101745 | 103680 | 6319 |
| # operations per optimization | 2 | 8.40 | 4871 | 16 |
| # operations per image | 12960 | 854658 | $5.05 \times 10^8$ | 101104 |
| Normalize values of the operations | 1 | 65.94 | 38968 | 7.80 |

TABLE III

MEASURING QUALITY OF THE PRESENTED METHODS THROUGH DIFFERENT SLIDING WINDOWS FOR FIG. 9

| Measuring Sliding Windows | 4×4 Non-overlapped | 4x4 overlapped | Gaussian overlapped | 8×8 Semi-overlapped |
|---|---|---|---|---|
| 4×4 Non-overlapped | 70.02 % | 63.49 % | 64.63 % | 63.06 % |
| 4×4 Overlapped | 67.68 % | 73.30 % | 73.35 % | 73.11 % |
| 11×11 Gaussian Overlapped | 84.10 % | 87.91 % | 87.94 % | 87.89 % |
| 8×8 Semi-overlapped | 81.62 % | 87.04 % | 87.45 % | 87.30 % |

Subjective appearance of watermarked image quality of various methods is shown in Figs. 9-11 and their precise SSIM scores for medium-detailed image (Fig 9)are tabulated in Table III . In this table each column represents the method used for subjectively optimized embedding the signature and each row shows the measured values of the entire image quality. Comparing the 4x4 non-overlapped with the overlapped method, it is seen while the non-overlapped image quality is optimum on its own, but when measured across the blocks its quality deteriorates. The quality of the semi-overlapped method is almost equal to the other two overlapped methods with no noticeable differences, but at a cost of much lower complexity as seen from Table II.

The similarity between evaluated SSIM indices of traditional 11×11 Gaussian overlapped and the semi-overlapped windows suggests that the semi-overlapped window can also be used instead of the traditional measure of image quality with a massive reduction in computational complexity. Also note that, as shown in Fig. 6 in low to medium watermarking strengths, the quality of non-overlapped method is very good and in these situations non-overlapped method with the least complexity can be used, without any side effects on image quality.

## I. Conclusion

This paper has shown that, through subjective optimization, the side effect of image watermarking can be greatly reduced. The DCT domain SSIM index was employed as a fitness function in quality optimization due to its practically high correlation with HVS. The SSIM based optimization process was first carried out on non-overlapped blocks of the original image. The quality of the watermarked image was very good in low to medium watermarking strengths and higher detailed images can stand larger watermarking strength while keeping visual quality. Since each block was optimized without any considerations to its neighbors, the disparity effect was pronounced in higher strengths. Optimization of overlapped blocks was introduced as a solution but its high computational complexity prevents using it in practice. To reduce the computational complexity, the optimization of semi-overlapped blocks was introduced. The semi-overlapped method not only preserved the original quality of the watermarked image as good as the overlapped methods but it also reduced the computational complexity significantly. The results also show the semi-overlapped window can be employed for measuring SSIM of pictures instead of traditional Gaussian overlapped windows, with a much reduced complexity.